\PassOptionsToPackage{numbers,sort&compress,square}{natbib}

\documentclass[pdflatex,sn-nature,iicol]{sn-jnl}

\RequirePackage{fix-cm}
\usepackage{graphicx}
\usepackage{anyfontsize}
\usepackage[numbers,square,comma,sort&compress]{natbib}
\usepackage{setspace}

\jyear{2025}
\begin{document}
\pagestyle{empty}

\title[When Excellence Stops Producing Knowledge]{When Excellence Stops Producing Knowledge: A Practitioner's Observation on Research Funding}

\author*[1,2]{Heimo Müller}\email{heimo.mueller@medunigraz.at}

\affil*[1]{Medical University Graz, Austria}
\affil[2]{Human Machine Mind Cooperation, Graz, Austria}

\abstract{
After almost four decades of participating in competitive research funding---as applicant, coordinator, evaluator, and panel member---I have come to see a structural paradox: many participants recognize that the current system is approaching its functional limits, yet most reform measures intensify rather than alleviate the underlying dynamics. This paper documents how excellence has become decoupled from knowledge production through an increasing coupling to representability under evaluation. The discussion focuses on two domains in which this is particularly visible: competitive basic research funding and large EU consortium projects. Three accelerating trends are examined: the professionalization of proposal writing through specialized consultants, the rise of AI-assisted applications, and an evaluator shortage that forces panels to rely on reviewers increasingly distant from the actual research domains. These observations are offered not as external critique but as an insider account, in the hope that naming a widely experienced but rarely articulated pattern may enable more constructive orientation.
}

\keywords{Research funding, Excellence, Evaluation, Goodhart's Law, Professionalization, AI-assisted proposals, Peer review crisis}

\maketitle
\singlespacing

\section{The Paradox: Everyone Knows, Yet Nothing Changes}

The central observation can be stated plainly: competitive research funding increasingly functions as a mechanism that optimizes for many things other than the production of new knowledge. This is not an accusation; it is a structural description.

Across roles---applicant, evaluator, program officer, and institutional administrator---a similar reality recurs: the system behaves like an optimization process that has drifted away from its original purpose. This drift is widely recognized and often discussed privately. Publicly, however, the system is treated as though it were functioning as intended.

The paradox deepens in the context of reform. Measures introduced in the name of quality, accountability, or fairness---more detailed evaluation criteria, stricter reporting requirements, additional transparency mechanisms, and increasingly professionalized support structures---tend to tighten the coupling between funding success and representational skill. As excellence is made more measurable, evaluation increasingly captures representability rather than epistemic contribution.

This is not primarily a failure of individuals. It is Goodhart's Law \cite{goodhart1975} operating at system scale: when a measure becomes a target, it ceases to be a good measure. In research funding, excellence becomes measurable by coupling it to anticipatable, representable, and evaluable features. These proxies---publication records, preliminary data, detailed work plans, institutional credentials, consortium composition---were intended as indicators of future knowledge production. Over repeated cycles, they become the object of optimization themselves. A closely related formulation is Campbell’s Law \citep{campbell1979}: the more a quantitative social indicator is used for decision-making, the more it becomes vulnerable to corruption pressures---and the more it distorts the very processes it is meant to monitor. In research funding, the indicator set is broader than a single metric (e.g., publication proxies, deliverables, formal completeness, ``impact'' narratives), but the mechanism is the same: once indicators determine resource allocation, they become targets of optimization in their own right \citep{campbell1979,goodhart1975}.

What makes this dynamic particularly insidious is that the system still appears productive by conventional measures. Papers are published, projects are completed, impact factors rise, and technology transfer occurs. The dysfunction is not a shortage of output but a drift in substance: funding increasingly favors research that can be specified with confidence in advance, while penalizing exploratory, uncertain, and foundational work from which unexpected discoveries often emerge.

These dynamics align with broader accounts of systemic flaws and incentive structures in contemporary science and its funding ecosystem \citep{alberts2014,edwards2017}, with formal analyses suggesting that contest-like funding competitions can be intrinsically inefficient \citep{gross2019}, and with evidence that national funding and evaluation regimes shape research-system performance \citep{vandenbesselaar2016}.

\section{Two Domains, Same Pattern}

The same coupling appears, in different forms, in competitive basic research funding and in large-scale consortium projects. Both matter because together they cover much of the contemporary funding landscape. Although examples in this paper are drawn from European programs, the structural dynamics are not EU-specific. Comparable patterns can be observed across national and international funding systems---from the NSF and NIH in the United States to UKRI in the United Kingdom, from the ERC to national research councils worldwide. The coupling between excellence and representability is a systemic feature of competitive, large-scale research funding, not an artifact of any single institutional arrangement.

\subsection{Basic Research: The Anticipation Trap}

In competitive basic research (national science foundations, research councils, excellence initiatives), the core problem is temporal: decisions must be made before knowledge exists. Evaluation therefore operates on representations---proposals that translate unknown futures into present legibility.

The consequences are predictable. Over funding cycles, certain representations succeed more reliably and become templates. Applicants learn to frame uncertainty conservatively, emphasize feasibility, present preliminary results, specify outcomes clearly, and align with current priorities. These are rational adaptations. Collectively, however, they shift the system toward research that can be credibly promised.

The tragedy is subtle. Risky work is not rejected through explicit policy; it is disadvantaged structurally. A proposal exploring genuinely open questions---where outcomes cannot be specified, where preliminary data may be misleading, and where the most honest answer is ``it is not yet known''---tends to score lower on accountability-oriented criteria. Risk mitigation becomes risk avoidance; exploratory work becomes incremental work.

Professionalization amplifies this effect. Institutions, recognizing that representational quality strongly shapes success, invest in grant-writing training, hire specialized support staff, and build internal review processes. These measures can be helpful. They also normalize an assumption that is rarely acknowledged openly: funding success depends substantially on proposal quality rather than research quality.

\subsection{Large-Scale Consortium Projects: Bureaucratization as Emergent Property}

Large consortium projects---multi-partner, multi-year, multi-million initiatives characteristic of programs like Horizon Europe---exhibit the same coupling, but with additional layers that make the structural dynamics more visible. Horizon Europe is explicitly organized as a framework programme with a defined pillar structure and implementation instruments (work programmes, clusters, missions, and European Partnerships, including institutionalised partnerships) \citep{heu_regulation_2021_695,heu_programme_guide,heu_partnerships}. Comparable ``big science by design'' architectures exist elsewhere: in the United States, NIH frequently uses cooperative agreement mechanisms to fund multi-component centres and consortia (e.g., U54 activity codes) and runs trans-NIH platforms such as the NIH Common Fund to catalyse cross-institute programmes \citep{nih_u54_activity,nih_commonfund_mission}. In the UK, UKRI similarly spans multiple councils and partnership-based strategic programmes \citep{ukri_explainer_royalsociety}.

The governance of such programs increasingly relies on implementation-centric models: specialized execution agencies, separated from research directorates, handle administration, evaluation, and monitoring. This separation is a rational response to scale, accountability demands, and administrative burden. The issue is not competence or intent. The issue is that this organizational solution tends to intensify the coupling between representability and success.

First, scale amplifies complexity. Multi-partner consortia, long timelines, and large budgets require extensive coordination before application. The successful proposal is not only a research document; it is also a coordination achievement. This creates a market for professional coordination services \citep{eurofreelancers,upgrants}.

Second, procedural requirements expand. Partner profiles, ethics reviews, data management plans, exploitation strategies, dissemination commitments, gender equality statements, open science compliance---each requirement serves a legitimate purpose. Collectively, they shift effort from research design toward administrative compliance.

Third, evaluation moves toward formalism. Under time pressure and high volume \citep{peer_review_professionalization,publons}, review tends to focus on what can be quickly assessed: formal completeness, consortial balance, alignment with call text, and clarity of deliverables. Deep epistemic judgment---the likelihood of producing important knowledge---becomes secondary to whether the proposal satisfies evaluative criteria.

A common system response is outsourcing. Proposal preparation increasingly migrates to specialized consultants who understand the evaluation system better than many researchers. Professionally written applications can outperform researcher-written ones \citep{innotrope}---not necessarily because the research is better, but because the representation is optimized for evaluation.

This produces a structural irony: execution agencies are meant to manage administrative burden that would otherwise overwhelm research evaluation. Yet separating administration from epistemic judgment does not dissolve the coupling; it formalizes it. The system becomes explicitly two-track: scientific content and administrative process, increasingly handled by different specialists. This is bureaucratization as an emergent property of accountability demands at scale.

\subsection{Frankenstein Consortia: When Integration Becomes Simulation}

Beyond a certain threshold, consortium size turns coordination into a representational exercise. This paper uses the term \emph{Frankenstein consortia} for collaborative structures in which formal integration increasingly substitutes for substantive integration.

A key distinction is scale \emph{with loose coupling} versus scale \emph{with enforced integration}. Communities, open networks, and knowledge ecosystems can grow large because participation is modular and selective: not everyone must synchronize with everyone else, and coherence can be maintained through shared norms, reusable artefacts, and voluntary coordination. Large consortia differ: they are contract-like structures expected to deliver integrated outputs on fixed timelines, remain auditable, and demonstrate coordination as an explicit deliverable. In this setting, adding partners does not merely add people; it multiplies dependencies and coordination load.

The communication bottleneck is mathematical. For $N$ partners, potential bilateral relationships scale as $N(N-1)/2$. Ten partners yield 45 channels; 30 yield 435; 50 yield 1{,}225. No consortium can sustain active, high-quality exchange across that space. Hierarchical structuring and work-package decomposition become necessary, but they shift integration toward vertical reporting rather than horizontal problem solving. In systems terms, scaling is managed by decomposing into \emph{nearly decomposable} subsystems---useful for manageability, but hostile to dense integration across the whole \citep{simon1962_architecture}.

Empirical work in science policy and organizational studies supports this mechanism. Coordination costs rise with distributed partnership structures and can erode outcomes as complexity increases \citep{cummings2007_coordination_costs_project_outcomes}. Reflections on European consortium-style funding similarly report that large schemes are often perceived as overambitious and difficult to coordinate in practice, with integration goals outpacing what consortia can reliably realize \citep{meirmans2025_transdisciplinary_funding}. At the same time, large international consortia can strengthen organizational capacity and research systems, yet this often comes with management overhead that can crowd out core scientific work if not carefully designed \citep{pulford2023_research_consortia_strengthen_systems}. Team size alone is also a weak predictor of high-value outcomes; collaboration \emph{architecture} often matters more than sheer scale \citep{zheng2025_synergy_not_size}. In other words, size can be carried by networks, but integration must be carried by structure---and structure has limits.

This motivates a threshold hypothesis (not a universal constant). Consortia with $N \approx 5$--$10$ partners can often maintain direct coordination. At $N \approx 10$--$20$, formal structures become necessary. At $N \approx 20$--$35$, direct integration becomes structurally difficult. Beyond $N > 35$, integration tends to exist primarily in documents rather than practice. Exact ranges vary with task interdependence, governance, and modularity, but the direction is robust: as the coordination surface grows, the marginal partner increasingly adds reporting and interface work rather than shared problem solving \citep{cummings2007_coordination_costs_project_outcomes,meirmans2025_transdisciplinary_funding}.

Coordination metrics make the pattern visible. Meeting time (person-hours/month normalized by project size), meeting density (meetings/partner/month), coordination ratio (coordination FTE / research FTE), and protocol volume (pages of minutes per research output) can serve as early warning indicators. When these increase faster than shared technical progress, coordination has decoupled from integration. The consortium remains active, but increasingly as paperwork: harmonized narratives, interface descriptions, and deliverables that certify unity without guaranteeing it.

AI-mediated coordination introduces a new risk. Large language models can generate coherent minutes, cross-work-package syntheses, and alignment narratives \citep{llm_grants_plos}. This can reduce administrative friction and help teams cope with scale. It can also lower the cost of producing the \emph{appearance} of coherence: divergent positions can be rendered compatible on paper, and unresolved disagreements can be documented as ``ongoing discussions.'' In a regime already optimized for representability, automated coherence makes simulation easier precisely when integration is hardest. Frankenstein consortia do not fail because participants are incapable; they fail because beyond a certain size, integration must be produced \emph{in practice} while the system rewards proofs of integration \emph{in artefacts}.

\section{Three Accelerating Trends}

The current moment is particularly urgent because three trends rapidly intensify the coupling between excellence and representability.

\subsection{Professional Proposal Writers: The Marketization of Access}

A substantial industry now exists for professional EU grant writing \citep{eurofreelancers,upgrants}. Some companies advertise success rates of 65\% for EIC Accelerator applications---roughly three times the average \citep{innotrope}. This is not fraud; it is expertise in navigating evaluation systems.

Systemically, this implies that access to funding increasingly depends on the resources required to hire professional writers. Well-resourced institutions and established networks can afford this. Smaller institutions, younger researchers, and unconventional projects often cannot.

The ethical problem is not whether individuals should use these services. Under competitive pressure, it can be rational to do so. The systemic question is different: when funding success depends more on professional representation than research quality, what kinds of research are most likely to be supported?

\subsection{AI-Assisted Proposals: Goodhart's Law Meets Language Models}

Large language models can now generate competitive grant proposals \citep{llm_grants_plos}. The American Heart Association allows AI use with disclosure. The National Science Foundation encourages disclosure while warning about falsification risk. The NIH does not ask who wrote applications.

This introduces a new dynamic. If representational polish becomes easier to produce, its value as a signal of competence declines, yet evaluation systems continue to reward it. The predictable result is an arms race in surface quality that further decouples representation from substance.

The deeper concern is not any single application but the emergent effect. When AI can generate plausible technical narratives, coherent work plans, and properly formatted consortia---requirements that once signaled serious preparation---what remains as a reliable quality signal? If AI assistance becomes universal, advantage may shift toward those with the best optimization practices rather than those with the best research ideas.

AI can reduce language barriers and writing-skill disadvantages, and in that sense it can democratize access. At the same time, it can standardize output and increase similarity across proposals, potentially biasing toward approaches that have textual precedent. In a system already inclined to favor representable work over exploratory work, AI assistance may accelerate that bias.

\subsection{The Evaluator Crisis: Quality Judgment by the Unqualified}

Perhaps most consequentially, an evaluator shortage appears structural rather than temporary.

Evaluation is unpaid or underpaid, time-consuming, and often receives minimal career recognition. Conscientious researchers carry a disproportionate reviewing burden while many contribute minimally \citep{pier2018}.

In large consortium calls, the issue intensifies through conflict-of-interest restrictions. In specialized fields, many of the most qualified reviewers have collaborative ties to likely applicants. Strict COI policies can therefore push panels toward reviewers who are distant from the relevant research domain.

Concerns about bias and conflicts of interest in peer review have a substantial empirical and conceptual literature \citep{peer_review_coi}.

A perverse dynamic follows: the more specialized and cutting-edge the research, the less likely that reviewers have deep domain expertise. Under those conditions, evaluation relies more heavily on surface signals---formal quality, consortial prestige, alignment with call language---precisely because substantive assessment requires expertise the panel may not have.

Agencies are aware of the problem and frequently struggle to recruit qualified reviewers. Common responses include relaxing expertise requirements, accepting reviewers from adjacent fields, and relying more on formal criteria. Each step can be rational in isolation. Collectively, these steps shift evaluation away from epistemic judgment and toward procedural assessment.

\subsection{The Self-Reinforcing Loop}

These trends reinforce each other.

Professional writers optimize proposals to match evaluation criteria, increasing the importance of form. This pushes others toward the same optimization, expanding the market. AI assistance accelerates production, lowering barriers while also standardizing output. Evaluators, overwhelmed by volume and constrained by COI, rely more on formal signals that professional writers and AI can optimize. This increases the value of professional optimization and completes the loop.

Each actor behaves rationally under constraints. Researchers hire writers to improve success probability. Writers optimize for evaluation because that is what they are paid to do. Evaluators use formal criteria because time and expertise are limited. Agencies increase procedural requirements because accountability demands it.

No single actor is at fault. The incentives are.

The resulting system resembles what organizational theorists call a ``competency trap'' \citep{levitt1988}: it becomes better at doing something that matters less. Representability is optimized because it can be measured. But representability is not knowledge production; it is the anticipation of knowledge production---and increasingly, it can become a simulation of anticipation.

\section{Why Reform Fails}

Standard reform proposals often miss the core issue. ``Better evaluation criteria'' typically make criteria more detailed and therefore reward more sophisticated optimization. Reviewer training improves consistency but does not question whether criteria capture epistemic value. Transparency makes optimization strategies easier to learn. Stricter ethics requirements add procedural layers and shift effort toward compliance. Expanded impact assessment creates new metrics to optimize.

Each reform assumes that the problem lies in implementation quality rather than in structural logic. Yet the logic itself is central: any system that allocates funding before knowledge exists will optimize for whatever proxies it uses. Making proxies more sophisticated does not break the coupling; it makes optimization more complex.

Reforms that could help would have to loosen the coupling: instruments that explicitly tolerate uncertainty for exploratory work; longer time horizons that reduce pressure for premature specification; evaluation that prioritizes question quality over anticipated answers; protected funding not subject to continuous competition; acceptance of negative results as legitimate outcomes; and coordination benchmarks that trigger review when thresholds indicate integration failure.

These reforms are difficult because they conflict with accountability logic: how can public funding be justified for research that may fail, may change direction, and may not deliver promised outcomes? The answer is epistemic: discovery requires uncertainty. The difficulty is that this answer is hard to operationalize in systems designed for predictability.

\section{A Personal Reflection}

After many years in this system---writing proposals, coordinating projects, evaluating applications, and serving on panels---the emotion that accumulates is not cynicism but sadness.

Examples recur. Brilliant researchers lose funding because proposals honestly acknowledge uncertainty. Mediocre projects are funded because work plans are detailed and deliverables are concrete. Panels spend more time on formal requirements than on scientific merit. Professional writers produce higher-scoring proposals than researchers can often produce on their own.

The system also works at times: funding is allocated to important questions, exploratory work is supported, and unexpected discoveries are enabled. The problem is not that the system is non-functional; it is that it is increasingly misaligned. It optimizes for the wrong objective.

What makes this especially frustrating is that the misalignment is widely recognized. In conference coffee breaks, in private conversations after panel meetings, and in late-night discussions among coordinators, the same observations recur: the system is becoming more formalized, bureaucratized, and professionalized in ways that select against the kind of research it was designed to support.

Publicly, however, the fiction is maintained. Proposals are written as if representational quality were epistemic quality. Criteria are applied as if they captured research value. Reports are produced as if completed deliverables necessarily implied knowledge produced.

Why does this persist? Because the system rewards conformity. Opting out can mean losing funding. Criticism without alternatives feels futile. Participation makes everyone, including the present author, complicit.

\section{Conclusion: Naming the Pattern}

This paper does not propose a complete solution. That choice is deliberate. Before alternatives can be evaluated productively, the current reality must be acknowledged clearly.

Competitive research funding has optimized itself into a state where excellence can mean ``excellent at getting funded'' rather than ``excellent at producing knowledge.'' This is not a collapse; it is an optimization toward the wrong target.

The opening paradox---everyone knows, yet nothing changes---reflects a deeper trap. Researchers require funding and therefore optimize for current criteria. Evaluators apply the criteria provided. Agencies respond to political accountability demands. Each role is rational within constraints. Collectively, the system reproduces the misalignment.

What could break the cycle? Explicit recognition that the current optimization has become dysfunctional. Not failed---dysfunctional. The system produces outcomes, but increasingly those outcomes are excellent proposals rather than excellent knowledge.

Artificial intelligence is likely to radicalize this tendency. If representational polish can be automated, proposal generation commoditized, and surface quality decoupled from effort and competence, the question becomes: what will evaluation optimize for next?

The purpose of this paper is not denunciation but articulation. Many researchers recognize fragments of this pattern; few have incentives to describe it publicly while still participating. Yet the pattern is real, intensifying, and it undermines the epistemic function that justifies research funding.

If the paper serves a purpose, it is to make the observation explicit: the system increasingly selects for excellence-as-representability. Whether it continues to select for excellence-as-discovery is an urgent question.

The answer is unlikely to come from ever more detailed criteria, more rigorous procedural layers, or enhanced accountability mechanisms. It will require acknowledging that some forms of knowledge production cannot be anticipated, specified, or evaluated in advance---and that funding systems must create space for uncertainty if discovery is to remain possible rather than merely simulated.

\subsection{Beyond Critique: Grounds for Optimism}

This analysis need not end in resignation.

Systems that cannot reform themselves from within eventually face external pressure. New movements emerge, alternative models develop, and calls for ``slow science'' argue for institutional time horizons and evaluative cultures that better accommodate uncertainty and depth \citep{slow_science}. What appears immutable can prove contingent.

Where money and power are directly involved---commercial funding, venture capital, private foundations---change can occur more easily because these actors can experiment with allocation mechanisms that prioritize performance over procedural defensibility. Some fund based on track record rather than detailed work plans; others support researchers for longer periods without demanding over-specified deliverables. These approaches are imperfect, but they demonstrate that alternatives exist.

For competitive public research funding---particularly large-scale strategic programs such as Horizon Europe---reform is harder. Political accountability, public transparency, and distributional fairness impose real constraints. Even so, the reform vocabulary is becoming more concrete.

A promising shift is the move from rhetorical calls for ``better peer review'' toward \emph{testable interventions in funding processes}. Recent work calls for empirical experiments on allocation mechanisms and decision reliability \citep{schweiger2024_costs_competition}. In parallel, funders and meta-research groups have begun to operationalize \emph{partial randomisation} and related experimental designs---not as ideology, but as a way to measure what current procedures actually achieve \citep{stafford2024_partial_randomisation,rori_experimental_funders_handbook}. Related proposals discuss how lotteries can be embedded in existing systems with different fairness and efficiency trade-offs \citep{feliciani2024_funding_lotteries}. The common point is not to abolish review, but to reduce overfitting to fragile proxies by introducing controlled uncertainty where fine-grained ranking is not reliable.

Optimism also comes from governance regimes that begin with articulated needs and target outcomes rather than maximal narrative completeness. Funding architectures exist that prioritize demonstrable capability development over exhaustive specification: problems are framed as concrete gaps, and evaluation focuses on credible delivery mechanisms rather than on narrative comprehensiveness. Such regimes can reduce the premium on prose-level representational perfection and increase the premium on tangible progress.

What might catalyze broader change? Possibly the dynamics described in this paper. If professional optimization makes proposals indistinguishable, AI assistance commoditizes surface quality, and reviewer shortages make deep assessment impractical, the contradictions become too visible to ignore. Crisis can create opportunity for reconstruction.

Resistance also exists within the system: researchers who refuse to optimize purely for fundability; program officers who recognize the gap between what is selected and what ultimately matters; evaluators who question whether rubrics capture the relevant signal. These actors are often isolated. Once dysfunction becomes undeniable, they may be able to form coalitions for change.

Funding systems are human constructions. They can be reconstructed. Accurate naming of the problem is not the endpoint; it is the beginning of informed change.

Research funding is justified by its connection to knowledge production. When that connection weakens, legitimacy erodes. Eventually, the question arises: why is this being done? When that question is asked seriously, alternatives become imaginable.

This paper describes a structural misalignment, not a destiny. Structures are choices that have solidified into habit. With sufficient recognition and will, different choices become possible.

The system will not fix itself. But it can be fixed. The first step is to see clearly what has been built.

\section*{Acknowledgments}

I am grateful to the many colleagues with whom I have collaborated over the years, to those encountered in review panels and evaluation committees, and to others who have shared the path through the research funding landscape. These experiences---often difficult, sometimes frustrating, occasionally inspiring---have shaped the observations presented here.

This manuscript was written with assistance from AI systems, which proved valuable for a non-native English speaker in formulating clear arguments and maintaining consistent academic prose. The irony of using AI support to criticize the role of AI in proposal production is fully acknowledged.

\bibliography{references}

\end{document}